\documentclass[12pt,preprint]{aastex}
\slugcomment{Accepted for publication in ApJL}

\def\xte{{\it RXTE}}

\def\Msun{\hbox{$\rm\thinspace M_{\odot}$}}

\begin{document}

\title {Frequency-dependent time lags in the X-ray emission of the Seyfert
galaxy NGC 7469}

\author {I. E. Papadakis\altaffilmark{1},
K. Nandra\altaffilmark{2, 3} and D. Kazanas\altaffilmark{4}
}

\altaffiltext{1}{Department of Physics, University
of Crete, 71003, Heraklion, Greece}
\altaffiltext{2}{Laboratory for High Energy Astrophysics, Code 662, 
	NASA/Goddard Space Flight Center,
  	Greenbelt, MD 20771}
\altaffiltext{3}{Universities Space Research Association}
\altaffiltext{4}{Laboratory for High Energy Astrophysics, Code 661, 
	NASA/Goddard Space Flight Center,
  	Greenbelt, MD 20771}

\begin{abstract}

We report the discovery of time lags in the cross-spectra of the X-ray
light curves of the Seyfert galaxy NGC 7469. This behavior is common in
Galactic black hole and neutron star binaries, and is in the sense that
harder X-rays are delayed with respect to the soft with a time
lag approximately proportional to the Fourier period. At the longest
period probed by our observation ($\sim 6$ days) we find a time lag of
approximately 3.5 hours between the 2-4 and 4-10 keV X-rays. A similar lag
and period dependence is found comparing the 2-4 and 10-15 keV light
curves, albeit with less significance. We find the coherence function of
the light curves to be close to 1 in the frequency range between
$10^{-5.5} - 10^{-3.5}$ Hz.  The implications of these results for the
X-ray production mechanism in active galactic nuclei (AGN) is discussed.

\end{abstract}


\section{INTRODUCTION}
\label{Sec:Introduction}

Both stellar mass and supermassive black holes show persistent, rapid,
large amplitude variations in their X-ray flux.  Analysis of this
variability can be a powerful diagnostic of the emission processes in
accreting compact objects. The variability is usually characterized by the
power spectral density (PSD) function which has a steep ``red noise'' form
that flattens below a characteristic break frequency. The dominant
radiation mechanism in these objects is thought to be inverse Compton
scattering of soft ``seed'' photons by a hot plasma (e.g. Shapiro,
Lightman \& Eardley 1976, Sunyaev \& Titarchuk 1980). A general prediction
of these models is that hard X-ray variations should lag those in softer
bands, as former photons undergo more scatterings in the Comptonizing
plasma.

Time lags have been observed in the Fourier cross-spectrum of both
persistent and transient galactic black hole candidates (GBHC; e.g.
Miyamoto et al.~1988, 1991). These observations lend strong support to
the Comptonization hypothesis as they indeed show that the soft X-ray
variations precede the hard. The observed time lag is an increasing
function of Fourier period, i.e. the more slowly varying components show a
longer time lag.  In the most famous case, Cyg X-1, the lags increase from
$10^{-4}$s on a time scale of 0.01s, flattening out to 0.1 above a
period of about 10s.

The variability properties of Galactic and extragalactic black holes are
quite similar.  For example, the form of the power spectra is very
similar, with both classes showing evidence for a low-frequency ``knee''
(e.g. Edelson \& Nandra 1999; McHardy et al.~1998) and a high frequency
``break'' (e.g. Nowak \& Chiang 2000). One AGN with a particularly
well-sampled X-ray light curve is NGC 7469 (Nandra et al. 1998). Nandra et
al. (2000) have shown that the X-ray spectral shape correlates with the
ultraviolet flux in this object. This is highly suggestive of the seed
photons cooling the Comptonizing plasma (e.g. Pietrini \& Krolik 1995;
Petrucci et al. 2000).  Nandra \& Papadakis (2001; hereafter NP01) found
further support for the Comptonization model in that the cross-correlation
functions (CCF) between X-ray light curves in various bands were
asymmetric towards positive lags (i.e. in the sense that the soft X-ray
variations preceded those in the harder bands). The variability amplitude
was found to be larger in the soft band also, at least on long time
scales.  Contrary to the Comptonization hypothesis, however, NP01 found
evidence that the high frequency PSD ``hardens'' (i.e. flattens) as a
function of energy. This behavior is also seen in Cyg X-1 (Nowak et al.
1999a), further extending the suite of similarities between AGN and GBHC.

Comparison between GBHCs and AGN has not been extended to the
cross-spectra, mostly because datasets of sufficiently high quality and
sampling have not been obtained for AGN. An exception is the NGC 7469
dataset just mentioned, and in this Letter we present an analysis of its
cross spectrum.

\section{ANALYSIS TECHNIQUES}

\subsection{Phase spectrum and coherence function}

Let $x_{s}(t)$ and $x_{h}(t)$ be the X-ray flux light curves in two energy
bands ($s$ and $h$ stand for ``soft'' and ``hard''), and let $R_{s}(k)$
and $R_{h}(k)$ denote the auto-covariance function of the two processes at
lag $k$ ($R(k)=E\{[x(t)-\bar{x}][x(t+k)-\bar{x}]\}$, where $E$ is the
expectation operator). The power spectral density of the processes is then
defined as $P(\omega)=(1/2\pi)\int R(k)e^{-ik\omega}dk$. The functions
$R_{s}(k), R_{h}(k)$, in the time domain, and $P_{s}(\omega),
P_{h}(\omega)$, in the frequency domain, characterize the variability
properties of the individual time series. If the two series are related,
the correlation structure between them is characterized by the
cross-covariance function, $R_{sh}(k)=
E\{[x_{s}(t)-\bar{x}_{s}][x_{h}(t+k)-\bar{x}_{h}]\}$ and, in the frequency
domain, by its Fourier transform, the cross-power spectral density
function (or simply cross-spectrum), $P_{sh}(\omega)=(1/2\pi)\int
R_{sh}(k)e^{-ik\omega }dk$. While $P(\omega)d\omega$ represents the
contribution to the time series variance of the components with frequency
$\omega$, $P_{sh}(\omega)$ represents the covariance between these
components in the two processes.
  
Unlike the PSD, the cross-spectrum is a complex function, since
$R_{sh}(k)$ need not be a symmetric function around $k=0$. Therefore, it
can be written as:
$P_{sh}(\omega)=c_{sh}(\omega)+iq_{sh}(\omega)=|a_{sh}(\omega)|
e^{i\phi_{sh}(\omega)}$. The cross-spectrum amplitude,
$|a_{sh}(\omega)|=\sqrt{c_{sh}(\omega)^{2}+q_{sh}(\omega)^{2}}$,
represents the average value of the product of the amplitudes of the
components with frequency $\omega$ in $x_{s}(t)$ and $x_{h}(t)$. The
function $\phi_{sh}(\omega)=tan^{-1}[q_{sh}(\omega)/c_{sh}(\omega) ]$ is
called the phase-spectrum and represents the average value of the phase
shift, $\phi_{h}(\omega)-\phi_{s}(\omega)$, between the components at
frequency $\omega$ in $x_{s}(t)$ and $x_{h}(t)$. The corresponding time
shift is given by $\tau(\omega)=\phi_{sh}(\omega)/\omega$. Finally, the
coherence function is defined as:
$\gamma^{2}(\omega)=|P_{sh}(\omega)|^{2}/[P_{s}(\omega)^{2}P_{h}(\omega)^{2}]$
(Vaughan \& Nowak 1997) and gives the fraction of the mean-squared
variability of one time series at $\omega$ that can be attributed to the
other (Nowak et al. 1999a).

\subsection{Estimation of lags and coherence function.}

Consider $N$ observations of the processes $x_{s,h}(t)$ for $t=1\Delta t,
2\Delta t, \ldots, N\Delta t$. The sample cross-spectrum between
$x_{s}(t_{i})$ and $x_{h}(t_{i})$ can be estimated by the cross
periodogram, $I_{sh}(f)$ ($f$ is the Fourier frequency, $f=\omega/2\pi)$:
$I_{sh}(f_{i}) = \{C_{s}(f_{i})C_{h}(f_{i}) + S_{s}(f_{i})S_{h}(f_{i})\} +
i\{ C_{s}(f_{i})S_{h}(f_{i})-S_{s}(f_{i})C_{h}(f_{i})\}$, where
$f_{i}=i/(N\Delta t), i=1,2,\ldots,N/2$, and $C(f_{i}), S(f_{i})$ are the
finite cosine and sine Fourier transforms of the data sets, i.e.
$C(f_{i})= \sqrt{\Delta t/N} \sum_{j}x(t_{j}) \cos (2\pi f_{i}) t_{j}$ and
$S(f_{i})= \sqrt{\Delta t/N} \sum_{j}x(t_{j}) \sin (2\pi f_{i}) t_{j}$.

In practice, it is customary to divide the light curve into many segments,
compute the cross periodogram for each of them, average
$Re\{I_{sh}(f_{i})\}$ and $Im\{I_{sh}(f_{i})\}$ (where $Re$ and $Im$
denote the real and imaginary parts) and/or rebin them by averaging say
$m$ consecutive frequency bins to obtain an estimate of the
cross-spectrum. In this case, the sample estimates of the phase spectrum
and coherence function are given by, $\hat{\phi}_{sh}(f_{i}) =
tan^{-1}(<Im\{I_{sh}(f_{i})\}>/<Re\{I_{sh}(f_{i})\}>)$ and
$\hat{\gamma}_{sh}(f_{i}) =
<|I_{sh}(f_{i})|^{2}>/<I_{s}(f_{i})><I_{h}(f_{i})>$, where
$|I_{sh}(f_{i})|^{2}=Re\{I_{sh}(f_{i})\}^{2} + Im\{I_{sh}(f_{i})\}^{2}$,
$I_{s}(f_{i})$ and $I_{h}(f_{i})$ are the periodograms of the individual
light curves, and the brackets denote an average over the ensemble of the
sample cross spectra and periodograms and/or over $m$ consecutive
frequency bins.

\subsection{The NGC~7469 light curves}

We have used the data from the $\sim 30$ day long \xte\ observation of
NGC~7469 in June/July 1996 in order to calculate the phase spectrum and
the coherence function between light curves at different energy bands. We
extracted light curves in the energy bands $2-4$, $4-10$ and $10-15$ keV
from the top layer of the proportional counter array (PCA). Full details
of the PCA data analysis are given in Nandra et al. (2000) and NP01. We
have used bin sizes of 5760s ($\sim$ the \xte\ orbital period) and 16s
for the analysis.

There are $N=480$ points in the 5760 sec binned light curves. Firstly, we
calculated the real and imaginary part of the cross-spectral function
between the $2-4/4-10$ keV and $2-4/10-15$ keV light curves. We also
calculated the sample power spectra of the individual light curves (see
section 5.1 in NP01).  The light curves are not long enough to divide them
into smaller segments so we simply grouped the real and imaginary parts of
the cross periodogram into bins of size $m$ and accepted them as our final
estimates of the cross-spectrum at the mean frequency of each bin. For the
$2-4/4-10$ keV cross spectrum, we chose a value of $m=10$ for the two
lowest frequency bins (in order to extend the sample cross-spectrum
estimation to the lowest possible frequencies) and $m=35$ for the
remaining bins. For the $2-4/10-15$ keV cross spectrum, we used a value of
$m=15$ for the two lowest frequency bins and $m=60$ for the higher
frequency bins. The larger $m$ value is necessary in this case as the
Poisson noise level is more prominent in the 10-15 keV band (see below).

In order to extend the phase spectrum and coherence function estimation at
the highest possible frequency we used the 230 parts of the 16 sec binned
light curves that have no gaps in them (see NP01) and estimated the sample
cross periodogram as before. We then combined all of them, sorted them in
order of increasing frequency and grouped them in bins of size 60 for both
cross spectra.

Poisson noise in the X-ray light curves affects the estimation of both the
phase spectrum and coherence function (for a detailed discussion see
Vaughan and Nowak 1997, Nowak et al. 1999a). For this reason, we corrected
for the Poisson noise power level the sample auto and cross-power spectra
used in the calculation of the coherence function as described by Vaughan
and Nowak (1997). We used their equation (8) to measure the uncertainty in
the estimates of the coherence function and equation (16) from Nowak et
al. 1999a in order to estimate the uncertainty in the sample phase
spectra.

\section{RESULTS}

\begin{figure}
\plottwo{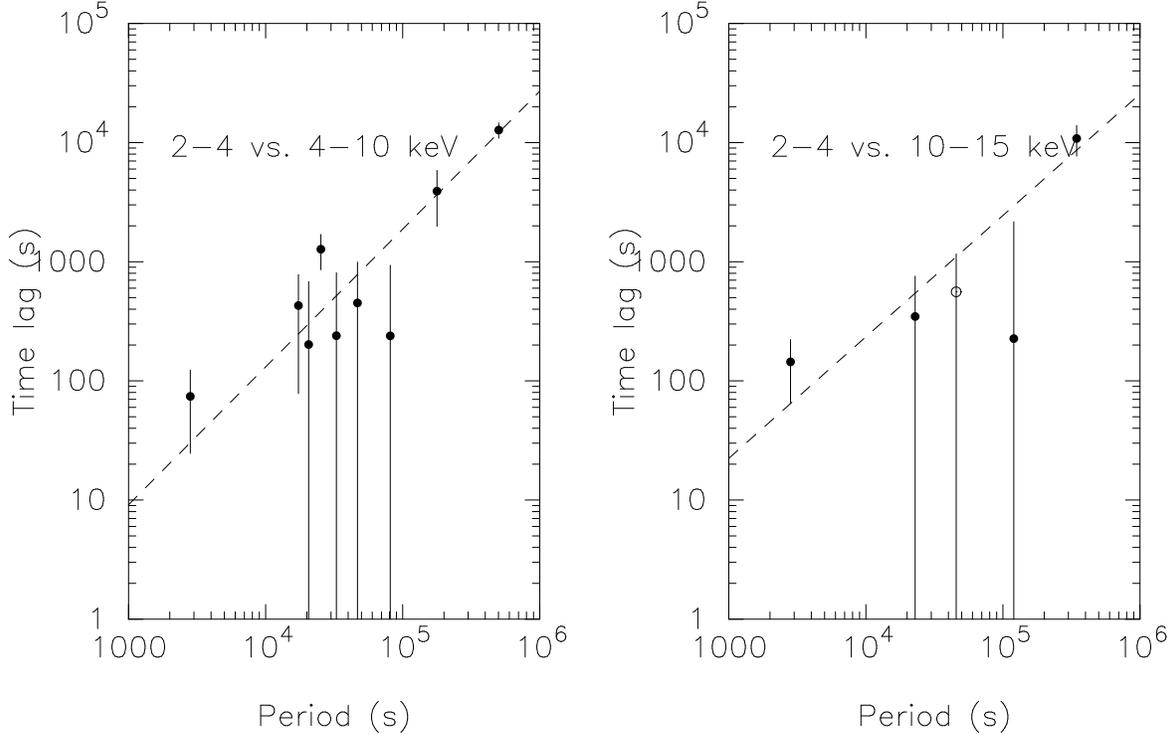}{fig_sh_cross.ps}
\caption{Time lag versus Fourier period ($1/f$) for the cross spectrum of
the soft (2-4 keV) versus medium (4-10 keV) band (left panel) and the soft
versus hard (10-15 keV) band (right panel). The lags have been binned as  
described in the text. The dashed lines show the best-fit power law model,
which gives a good fit. In the former case this shows a slope of $1.2\pm  
0.2$ (68 per cent confidence for one interesting parameter) and
normalization of $2.6 \times 10^{-3}$. The latter is much less well
determined due to the higher Poisson noise level in the 10-15 keV band.  
Indeed the point shown with the open circle has a negative lag. Excluding
that from the fit we find a best fit of 1.0 with a 90 per cent lower limit
of 0.6 and no upper bound.}
\end{figure}

Our results for the $2-4/4-10$ and $2-4/10-15$ keV light curves are
shown in Fig. 1 and 2. Fig. 1 shows the ``time lags" (i.e. the phase
spectrum divided by $2\pi f$) as a function of the Fourier period
(i.e. $1/f)$. The time lags for the $2-4/4-10$ keV light curves are
all positive and increase with period. The positive value indicates
that components at frequency $f$ in the $4-10$ keV band are delayed
with respect to the same components in the $2-4$ keV band. The amount
of the delay does not remain constant but increases with period. A
constant lag does not give a good fit to the data ($\chi^{2}=58.8/8$
d.o.f).  For the components at period $\sim 3$ ksec the delay is $\sim
100$ sec, increasing to $\sim 10$ ksec for components at period $\sim
500$ ksec ($\sim$ 6 days).  We observe very similar behavior for the
time lags between the $2-4$ and $10-15$ keV light curves.  The
determination of the lags is less precise in this case, however, and
in fact one of the points in the $2-4/10-15$ keV time lags plot is
negative.  When calculated using the 16sec binned light curves, the
errors in the phase spectra are large at all but the lowest frequency
point, and only this is plotted in Fig.~1.  We have fitted a power law
model to the time lag plots, with a slope of $\sim 1$ fitting both
datasets well.

Fig. 2 shows the coherence functions. Although the uncertainties are
rather large at high frequencies, our results suggest that intrinsic
coherence is roughly unity for both the $2-4/4-10$ and $2-4/10-15$ keV
bands over $\sim$ two decades of frequencies.

\begin{figure}
\plottwo{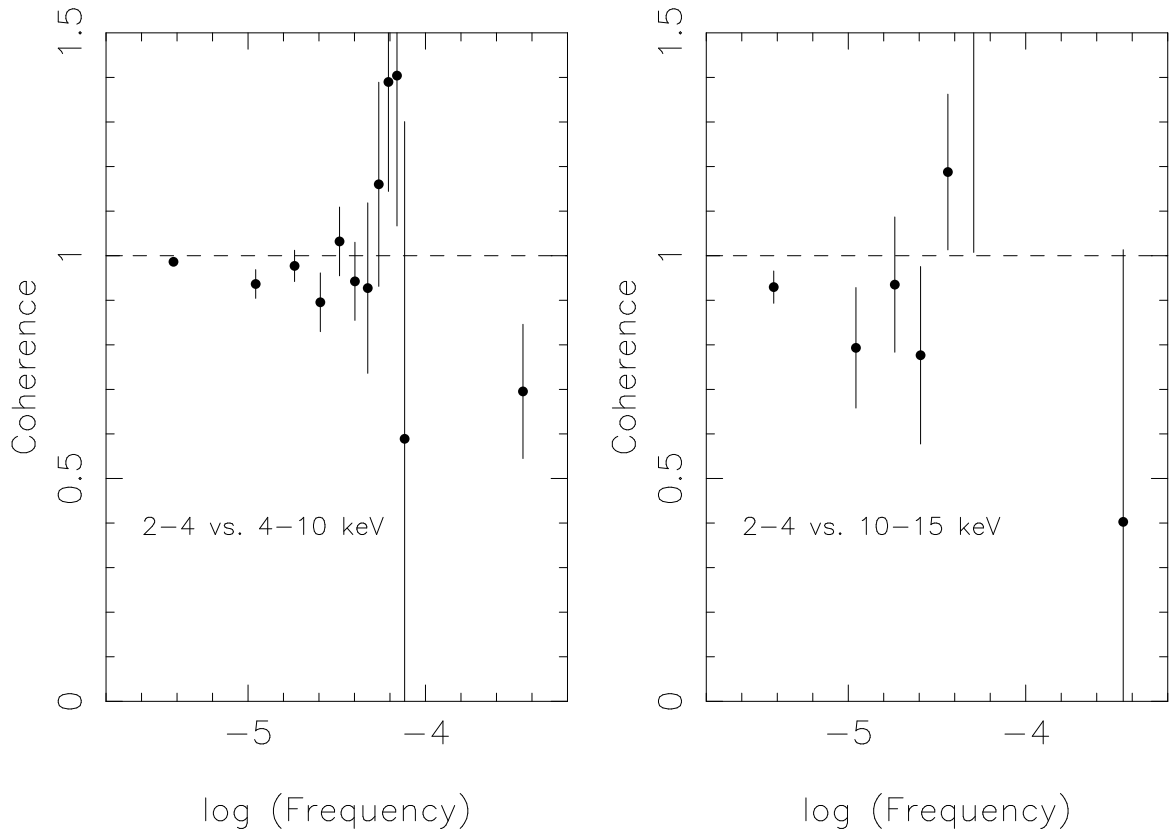}{fig_sh_coh.ps}
\caption{Coherence Fourier frequency soft versus medium band (left panel)
and the soft versus hard band (right panel). The coherence is binned in  
the same way as the lags in Fig.~\ref{fig:cross}. The values are close to
or consistent with 1 (dashed line) similar to Cyg X-1.
\label{fig:coh}}
\end{figure}

\subsection{Simulations}

We have performed light curve simulations to validate the significance of
the time lag detection. Our first set of simulations investigates the
effects of Poisson noise which contributes a random component to the
measured phase lag at each frequency. While this introduces an
uncertainty, we do not expect that it will result in systematic lags
between two time series, as we observe.  We used the observed $2-4$ keV,
5760s light curve as a template, and created 1000 pairs of light curves by
adding random noise to each point equal to the average Poisson error of
the observed light curves. The pairs were therefore identical (and hence
have a coherence unity by construction), except for the effects of the
Poisson noise and have no phase lags at any frequency.  We computed the
phase spectrum for each pair of light curves in the same way that we did
for the actual data. The average lags were always very close to zero.
Considering the two lowest frequency points in Fig~1a, the observed value
was never achieved in the 1000 trials (also true of the point at period
$\sim 25$ ksec).  Furthermore, no phase spectrum of the 1000 synthetic
light curves had all the estimated time lags being positive as we observe.

We then repeated this exercise using a randomly-generated light curve,
rather than the real one, to investigate the effects of the red-noise
character of the source variations. We assumed a PSD with a slope 1.5
appropriate for the 2-4 and 4-10 keV light curves (NP01), and created 1000
pairs. As before the pairs were identical except for differing Poisson
noise.  Our results were very similar as in the previous case. The two
lowest frequency points in Fig. 1a have values larger than the time lags
of all 1000 simulated light curves, and none of the 1000 phase spectra had
all the points positive. We conclude that detection of lags in the phase
spectrum shown in Fig. 1a cannot cannot be the result of random, Poisson
or red noise fluctuations in the $2-4/4-10$ keV light curves.

Having established that the phase lags are not an artifact of Poisson or
red noise in the data, we finally attempted to test whether the error
prescription of Nowak et al. (1999a) is appropriate for our data. Although
the dispersion of the points in the above simulations was found to be very
similar to the calculated errors, a fairer test would be to simulate light
curves with similar time lags to those observed in our data. We achieved
this by performing a simulation similar to that just described, but
shifted the phase of the sinusoids in the second series by the best-fit
value obtained for the power law fit to the data in Fig.~1a. Random noise
was then added to both series. Here we easily recovered the time lags from
the simulated light curve: a positive lag was found for all 1000
simulations in the two lowest frequency bins. Furthermore, the dispersion
on the simulated legs was very similar to the errors prescription of Nowak
et al. (1999a), validating that technique for our data set.

\section{DISCUSSION}

We have presented the first cross-spectral analysis of the X-ray emission
of a Seyfert galaxy, NGC 7469. We find time lags in the sense that the
hard X-ray emission is delayed with respect to the soft. The estimated
phase spectra have a rather large uncertainty, nevertheless the
observed lags are consistent with being proportional to the variability
time scale with a magnitude that is approximately 1 per cent of the
Fourier period. The detection of frequency dependent time lags is
consistent with the results from the cross-correlation analysis of the
X-ray light curves of NGC~7469 (NP01). The CCF shows no peak at time
different than zero lag and it is asymmetric, in the sense that the soft
X-rays lead the hard by a fraction of a day. The sense and magnitude of
this asymmetry is consistent with the time lags we observe in the cross
spectrum: although the peak of the CCF is dominated by the fastest
variations, the asymmetry is caused by the relative delays between the two
bands in the longer time scales. Finally, we find that the emission in the
2-4, 4-10 and 10-15 keV bands is highly coherent. 

An obvious conclusion from Fig. 1 is that the lags in NGC 7469
resemble very closely those in the best studied galactic X-ray source,
namely Cyg X-1 in its hard state (Miyamoto et al. 1992), but scaled by
a factor $\sim 10^{6}$. Indeed, both the dependence of the lags on the
Fourier period and their magnitude as a fraction of the given period
($\simeq 10^{-2}$) are very similar to those observed both in GBHC
(Hua, Kazanas \& Cui 1999) and neutron stars (Ford et al 1999).

The presence of lags in the sense we observe in our light curves offers
strong additional support to the hypothesis that the X-ray are most likely
due to Comptonization of soft photons by hot, thermal electrons. Such a
hypothesis is consistent with both the X--ray spectra (Gondek et al. 1996)
and the relationship between the seed photon flux and X-ray spectral shape
(Nandra et al. 2000; Petrucci et al. 2000). In addition, the magnitude and
period dependence of the lags places constraints on the processes
responsible for the emission of radiation and the associated variability.
There are several different models which account for these effects, and
these can be divided into two broad classes:

(a) Models in which the lags result from the excess time the hard photons
spend in the Comptonizing electron cloud before they escape (Kazanas, Hua
\& Titarchuk 1997; Hua, Kazanas \& Titarchuk 1997). Because the
propagation speed is that of light, the observed lags afford an estimate
of the size the X-ray source. The longest observed lag in our dataset is
$\sim 12,000s$ and Collier et al. (1998) estimated the mass of the black
hole in NGC 7469 to be $8 \times 10^{6}$~\Msun.  These models therefore
imply a size for the region of at least $240 M^{-1}_{\rm 7} \ R_{\rm g}$,
where $M_{7}$ is the black hole mass in units of $10^{7}$~\Msun and
$R_{\rm g}$ the gravitational radius.  Our dataset does not sample periods
longer than $\sim 10^{6}$s, but if the trend of lag increasing with period
continues, the implied radius would be even larger. Having such an
extended Comptonizing region raises the question of the plasma heating
process. ADAF models (e.g. Narayan \& Yi 1994) appear to provide a natural
mechanism, but they also predict the dependence of the lags on the Fourier
period to the radial density profile of the Comptonizing electrons.  The
linear dependence of the lag on the Fourier period we observe argues for
an electron density profile $n(r) \propto r^{-1}$. This profile is not
consistent with the ADAF solution, but it is consistent with its variant
ADIOS (Blandford \& Begelman 1999). The models of this class, as presented
to date, have employed an electron temperature which is independent of
both radius and time. However, the electron temperature is expected to
decrease within the ADAF models at a distance larger than $\sim 10^3 ~R_g$
while photon cooling should also contribute to the time dependence of the
electron temperature. Indeed, the inclusion of time-dependent Compton
cooling leads to significant modifications of the results compared to a
static-corona case (Malzac \& Jourdain 2000). Recently, B\"ottcher (2001)
has shown that a model with a slab corona geometry (e.g. Haardt \&
Maraschi 1991) and a small-scale flare of the underlying, optically thick
accretion disk, can explain well the overall shape of the time lag spectra
in GBHC without a large corona when the effects of cooling due to the soft
photon input is taken into account.

(b) Models in which the lags are produced by a spectral evolution of 
the X-ray emission (from soft to hard) due to coherent changes in the
emission region (Poutanen \& Fabian 1999; B\"ottcher \& Liang 1999;
Nowak et al. 1999b).  The magnitude and sign of the lags in this class
of models are obtained by a systematic hardening of the X-ray emission
with time. The observed lags then, along with an estimate of the
characteristic size of emission (a few tens of Schwarzschild radii),
leads to a characteristic speed for effecting the change responsible
for the observed X-ray variability. These speeds are generally much
lower than those generally associated with dynamical and thermal time
scales in the corresponding radii.

In conclusion, the time lags in the X-ray light curve of NGC 7469 are very
similar to those observed in the X-ray light curves of GBHC. This - along
with the corresponding PSDs - provides further evidence that the processes
which determine the dynamics of accretion and the emission of X-ray
radiation scale roughly linearly with the mass of the accreting object. 

\acknowledgements 
We thank the \xte\ team for their operation of the satellite. IEP
thanks the LHEA for their hospitality, and USRA for financial
support. KN is supported by NASA grant NAG5-7067 provided through the
Universities Space Research Association.

\clearpage


\end{document}